\documentclass[twocolumn,showpacs,preprintnumbers,amsmath,amssymb]{revtex4}

\usepackage{graphicx}% Include figure files
\usepackage{dcolumn}% Align table columns on decimal point
\usepackage{bm}% bold math

\begin{document}

\preprint{APS/123-QED}

\title{Ratios in Higher Order Statistics (RHOS) values of Seismograms for Improved Automatic P-Phase Arrival Detection }
\author{Mulugeta Dugda}
\email{mp.tuji@gmail.com}

\affiliation{Department of Electrical and Computer Engineering}
\author{Abebe Kebede }
%\email{abkebede@gmail.com}
\affiliation{Department of Physics \\ North Carolina Agricultural
and Technical State University\\ 1601 East Market Street,
Greensboro, NC 27411}
\date{\today}
\begin{abstract}
In this paper we present two new procedures for automatic detection
and picking of P-wave arrivals. The first involves the application
of kurtosis and skewness on the vector magnitude of three component
seismograms. Customarily, P-wave arrival detection techniques use
vertical component seismogram which is appropriate only for
teleseismic events. The inherent weakness of those methods stems
from the fact that the energy from P-wave is distributed among
horizontal and vertical recording channels.  Our procedure, however,
uses the vector magnitude which accommodates all components. The
results show that this procedure would be useful for
detecting/picking of P-arrivals from local and regional earthquakes
and man-made explosions. The second procedure introduces a new
method called "Ratios in Higher Order Statistics (RHOS)." Unlike
commonly used techniques that involve derivatives, this technique
employs ratios of adjacent kurtosis and skewness values to improve
the accuracy of the detection of the P onset. RHOS can be applied
independently on vertical component seismogram as well as the vector
magnitude for improved detection of P-wave arrivals.

\end{abstract}


\maketitle
\section{INTRODUCTION}
Detection of P-wave arrivals accurately is an important step to make
earthquake forecast, to analyze the interior structure of the earth
and to study earthquake sources. Picking seismic phase arrivals
correctly is also helpful to discriminate between natural
earthquakes and man-made explosions. Several P-wave detection
techniques utilize only one component of the usually recorded
three-component seismogram \cite{REF1}. These techniques work very
well for the case of teleseismic events (with epicentral distance
$\Delta\geq$ 3300 km) since seismic P-waves approach the seismic
sensors in a nearly vertical incidence (where $\Delta$ is the
epicentral distance between a seismic station and the seismic
epicenter) (Figure 2).

Magotra \cite{REF2} attempted to use horizontal component
seismograms in order to determine the direction of arrival of
seismic waves towards a seismic station. \cite{REF3} used kurtosis
and skewness of the vertical component seismogram to estimate the
P-onset time assuming that the maximum of the derivatives of the
kurtosis and skewness curve, just before the curve attains its
maximum, is considered to be the time of arrival for the P-wave.
Unlike in teleseismic events, the P-wave signal originating from
regional (100 km $\leq\Delta\leq$ 1400 km) and local distances
$\Delta <$ 200 km) arrive at the seismic station with significant
strengths in both horizontal and vertical directions
\cite{REF2}-\cite{REF4}. Since the energy is distributed between the
horizontal and the vertical recording channels, both horizontal and
vertical component seismograms need to be taken into account for
improved detection. In this study the seismic vector magnitude of
three component seismograms from a single station is used to
determine the P-wave arrival without employing the derivatives of
the kurtosis and skewness.

When a P-wave arrives, we can observe a maximum asymmetry of
distribution of the vector magnitude of the three-component
seismograms. As a result of this asymmetry, maximum values of
kurtosis and skewness are expected to occur when a P-wave arrives
and this information can be extracted and used for the P-arrival
detector. Thus, one essential hypothesis of this study is that the
normalized vector magnitude will show very high kurtosis and
skewness magnitude when a P-wave arrives. First we perform
window-by-window normalization of the vector magnitude to reduce the
huge variations in magnitude resulting from the ground motion and
obtain a zero-mean normalized vector magnitude for each window. The
normalized vector magnitude is then used as input to the detection
and picking system. The automatic detection and picking system makes
use of the skewness and kurtosis of the input normalized vector
magnitude in order to detect the P-arrival. For a sliding time
window of the seismic vector magnitude the values of kurtosis and
skewness are calculated and the time at which these values become
maximum/minimum is used to estimate the time of arrival of the
P-wave.

Another important contribution of this study is that unlike some
previous studies which are applied only on the vertical component
seismogram and used the derivatives within the kurtosis and skewness
values for their correction \cite{REF3}, our technique uses the
ratios within the kurtosis and skewness values rather than the
derivative to make a correction in the P-onset time. This new
technique supposes that the time for P-onset occurs when the HOS
(skewness and kurtosis) values of the vector magnitude begins to
change drastically during the course of gaining their maximum
values. In the next subsections we discuss the mathematical
background of the implemented technique, our proposed methods,
results, discussion, and some concluding remarks.

\section{Rationale and Mathematical Background}
\subsection{Background for Vector Magnitude and Normalization}

Broadband seismograms represent either the ground velocity or
displacement. If the three component seismograms representing the
motion of a seismic sensor in the three perpendicular directions
have magnitudes x, y and z, where x and y are usually the east-west
and north-south components, and z is the vertical component, the
magnitude of the resultant velocity/displacement vector v (Figure 1)
can be given by

\begin{equation}\label{eq:susceptibility2}
\\v =  \sqrt{\ x^2 + \ y^2 + \ z^2}.
\end{equation}

\begin{figure}[htb]
\center{\includegraphics[scale=0.4]{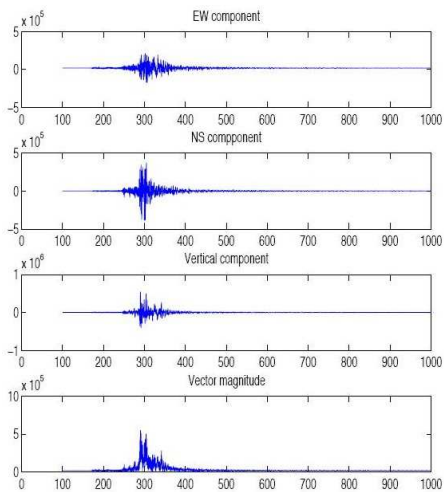}} \caption{Three
component seismograms (top three panels) and their vector magnitude
v(n) (the bottom panel) as an example.} \label{Fig}
\end{figure}

\begin{figure}[htb]
\center{\includegraphics[scale=0.5]{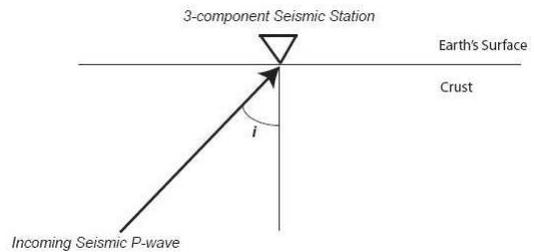}} \caption{Angle of
incidence i for an incoming P-wave approaching the seimic sensor of
a 3-component seismic station.} \label{Fig}
\end{figure}

For a sequence of digital seismograms, v can be calculated
sample-by-sample as

\begin{equation}\label{eq:susceptibility2}
\ v(\\n) =  \sqrt{\\x^2(\\n) + \\y^2(\\n) + \\z^2(\\n)}.
\end{equation}

where n is the nth sample. The angle of incidence for the seismic
ray in Figure 2 can be given in terms of x, y, and z component
values as

\begin{equation}\label{eq:susceptibility2}
i = \tan^{-1}\left(\frac{{\sqrt{\\x^2 + \\y^2}}}{z}\right).
\end{equation}

where $\sqrt{x^2 + y^2}$ represents the total horizontal component
of the seismic P-wave velocity. For teleseismic earthquakes
(events), the P-wave arrives the seismic sensor nearly vertically
and thus this angle of incidence is very small. For local and
regional earthquakes this angle of incidence becomes significant.
The method developed here is an improved version of the P Arrival
Identification-Skewness/Kurtosis technique\cite{REF3}. That method
is based on the observation that the noise in the vertical component
seismogram shows zero-mean Gaussian behavior until the P-wave
arrives. On the other hand, unlike the single components, the
seismic vector magnitude, v(n) is always greater than or equal to
zero, and it may not show zero mean Gaussian behavior. But we can
normalize this vector magnitude on a window-by-window basis not only
to have a zero mean Gaussian behavior, but also to better look at
the differences in the computed skewness and kurtosis. The
normalized vector magnitude is obtained through the following
transformation:

\begin{equation}\label{eq:susceptibility2}
\tilde{\\v_i}(\\n) =  \frac{\\v_i(\\n) - \bar{\\v_i}}{\sigma_i}.
\end{equation}

\begin{figure}[htb]
\center{\includegraphics[scale=0.5]{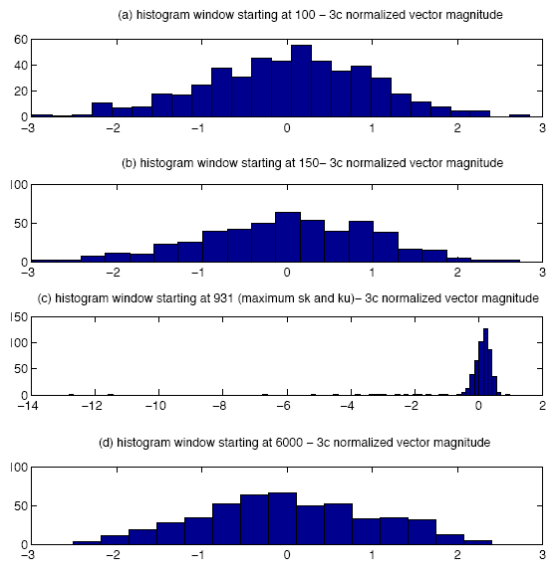}} \caption{A
histogram of normalized vector magnitude of 3-component seismograms
(for 1994 Rukwa seismic event of Tanzania recorded by temporary
seismic station Hale in Tanzania, East Africa). A window size of 500
samples is used here. (a) and (b) are for pure background noise,
window starting at sample 100 and 150 samples, (c) is when P-wave
arrives, and so it includes a lot of background noise and P-arrival
signal around the end of the 500 samples window starting at 931
sample, and (d) is 5000 samples after P-arrival. The skewness and
kurtosis values for these four windows are as follows: (a) skewness
= -0.1685, kurtosis = -0.0987 (b) skewness = -0.2278, kurtosis =
-0.1815 (c) skewness = -8.7562, kurtosis = 93.8140 (d) skewness =
0.0350, kurtosis = -0.5883.} \label{Fig}
\end{figure}

Each normalized variable $\tilde{v_i}(n)$ is a rescaled
(transformed) variable of sample $v_i(n)$ for an ith window of mean
value $\bar{v_i}$ and standard deviation $\sigma_i$. This
transformation is linear \cite{REF5}. Figure 3 displays the Gaussian
behavior of the normalized vector magnitude for a 3-component
seismogram. It also shows how the distribution and skewness and
kurtosis values for the normalized vector magnitude change
drastically when P arrives (Figure 3(c)). As Figure 3 clearly
displays, skewness and kurtosis receive much higher values when a
P-wave arrival is included in their calculation (Figure 3(c)) as
compared to their near-zero values when P-arrival is not included in
the sliding window (Figure 3(a),(b), and (d)).

Figure 4 depicts clearly that the kurtosis and skewness values take
their peaks just after P-arrives. This normalization process seems
to lead to enhancement of differences and sets the limit in the
range of variations. Figure 5 shows the enhancement in the maximum
values of the quantities under investigation when we apply
normalization as compared to the values without normalization.

\begin{figure}[htb]
\center{\includegraphics[scale=0.5]{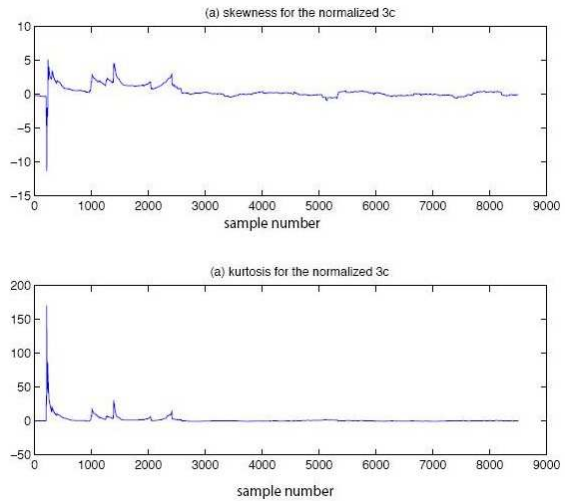}} \caption{(a) Skewness
and (b) Kurtosis values for the normallized vector magnitude of
3-component seismograms.} \label{Fig}
\end{figure}

\begin{figure}[htb]
\center{\includegraphics[scale=0.5]{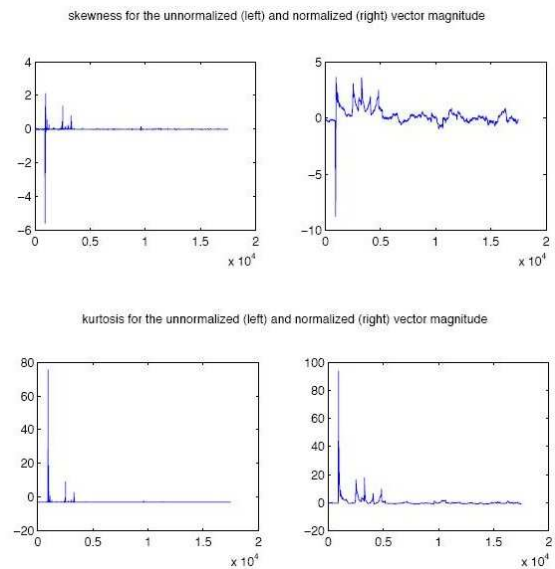}} \caption{Skewness and
Kurtosis of normalized and unnormalized vector magnitude of
3-component seismograms. All horizontal axes represent sample
number.} \label{Fig}
\end{figure}

\subsection{Mathematical Background for Higher Order Statistics}

Skewness(sk) and kurtosis (ku) for a finite length sequence v(n) are
estimated by:

\begin{equation}\label{eq:susceptibility2}
sk = \frac{1}{(K-1)\,\hat{\sigma}_s^3} \sum_{n=1}^K\,\left(v(n) -
\hat{m}_s\right)^3
\end{equation}

and

\begin{equation}\label{eq:susceptibility2}
ku  = -3 + \frac{1}{(K-1)\,\hat{\sigma}_s^4}{\sum_{n=1}^K\,(\\v(\\n)  \\
\\~- \hat{\\m_s})^4}
\end{equation}

Here $\hat{m}_s$ and $\hat{\sigma}_s$ are the mean and standard
deviation estimates of $v(n)$ and $K$ is the length of the finite
sequence. Generally, HOS values for a Gaussian distribution are
zero. In this case, it is demonstrated that the normalized magnitude
yields higher HOS values for P-arrival than the values calculated
for the background noise and P-wave coda (Figure 3(c)).

\section{Hypotheses and Procedure}

A generalized hypothesis applicable to local and regional events is
proposed. This hypothesis is based on the combination of all
three-component seismograms in a single station, and it is supposed
to outperform those techniques that are based on just a single
component seismogram. Following the mathematical equations in
section 2, the vector magnitude of three component seismograms for a
single station was calculated.  This is followed by normalization of
N-sample window of the vector magnitude (Eq. 4). While this window
slides to the right by one-sample at a time the window next to it is
automatically normalized. This normalization continues until the
last window is normalized. The skewness and kurtosis of each
normalized window were computed.

Pure background noise and the seismic signal away from P-phase
arrival follow a Gaussian distribution with nearly zero HOS values.
Because of high asymmetry and non-Gaussian distribution introduced
by the P-arrival on the Gaussian background noise, the window with
P-arrival follows a highly skewed distribution. Maximum (Minimum)
HOS values are attained for the window that includes the arrival of
the P-onset and just few additional samples of the P-wave coda.
Thus, to determine the P-onset time more accurately, a correction
scheme is introduced for the additional samples included after the
actual P-onset.

The PAI-S/K method of \cite{REF3}, using the vertical component
alone, proposes to use the location of the maximum slope as the
P-onset time. In this study, however, the P-onset time is taken as
the time when the HOS values of the normalized vector magnitude
start to increase sharply. The HOS values change drastically when P
breaks out from the background noise level during the P-onset. Thus,
the magnitude of the ratio of the skewness value on the
right-hand-side at the P-onset to the skewness value on the
left-hand-side (background noise) should be the maximum of all the
ratios within the window. The same is true with the ratio of
kurtosis values (Figures 6 to 9).
\begin{figure}
\center{\includegraphics[scale=0.5]{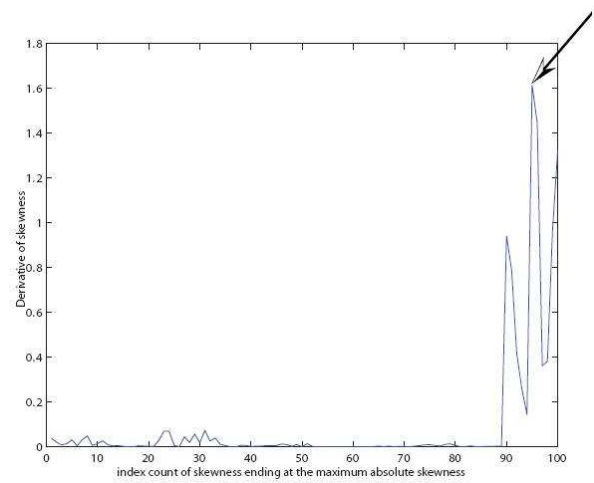}} \caption{
Application of maximum derivative of skewness for correcting
P-arrival time according to a 2002 study; it provides a correction
of 5 samples for the P-arrival.}
\end{figure}

\begin{figure}
\center{\includegraphics[scale=0.5]{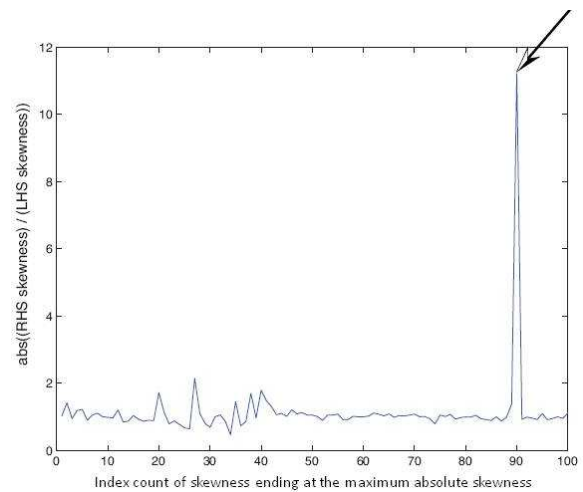}} \caption{
According to this new study, RHOS value of skewness gives a
correction of 11 samples for the P-arrival, which is by far better
than the 2002 study (Figure 6). The correct P-arrival happens 12
samples away from the sample of the maximum skewness, as shown in
Figure 10.}
\end{figure}

\begin{figure}
\center{\includegraphics[scale=0.5]{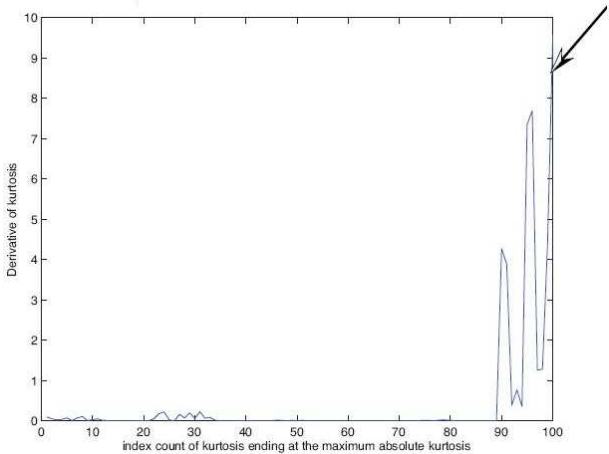}} \caption{Using
maximum derivative in kurtosis, according to a 2002 study, gives a
correction of 1 sample for the P-arrival.}
\end{figure}

\begin{figure}
\center{\includegraphics[scale=0.5]{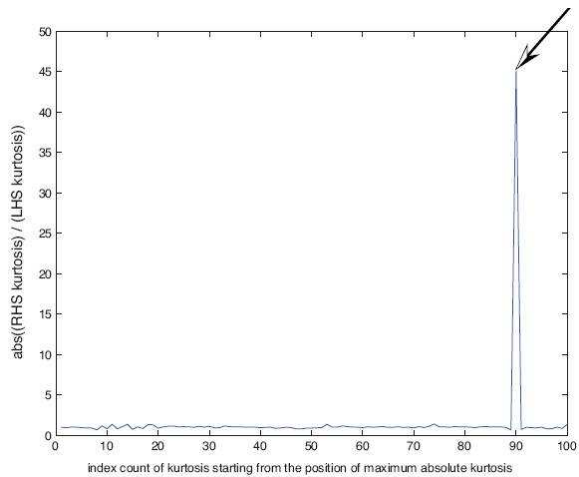}} \caption{
According to this new study, applying RHOS value of kurtosis offers
a correction of 11 samples for the P-arrival, which is consistent
with the skewness arrival coorection scheme of Figure 7, and a much
better correction than the 2002 correction (Figure 8); Moreover, the
corrections based on maximum derivatives of kurtosis and skewness in
the 2002 study do not agree with each other (Figures 6 and 8), while
the corrections by RHOS values of this study agree very well
(Figures 7 and 9).}
\end{figure}

\section {Results and Discussion}
\subsection{Experimental Setup}
For this particular study, we used seismic data sets obtained from
Integrated Research Institutions for Seismology (IRIS) Data
Management Center (DMC) depository website. Events from several
geological terrains were selected. The seismic networks contributing
to the IRIS/DMC which are utilized in this study include the
IRIS-PASSCAL Tanzania Broadband Experiment (1994/1995), the Global
Seismic Network (GSN), IRIS/USGS network, PFO of the IRIS/IDA,
Southern California Seismic Network (SCSN), and Pacific Northwest
Seismograph Network (PNSN).
\subsection{Results of the new scheme}
 Events within regional and local distance range of 50 km to 1400 km were
selected. Only broadband seismic data are used in this study.
Seismic records with different levels of noise were included. The
signal analysis was performed using Matlab 7.6, R2008a (3 GHz
Dual-Core Intel mac pro). The performance of the method using the
seismogram vector magnitude has been compared to the PAI-S/K
(Figures 6 to 9). The improvement of the arrival detection when
applying the ratios of HOS values as compared to using just the HOS
values directly as in \cite{REF3} is clearly shown for both skewness
and kurtosis. Further discussion is given in the next subsection.
Sizes of sliding window ranging from 30 to 500 samples are found to
exhibit best results.
\subsection{Discussion}

Comparison with PAI-S/K and other schemes has been made. There are
three important differences between the new technique presented in
this paper and PAI-S/K \cite{REF3} method. The PAI-S/K method is
among the many vertical component utilizing methods while our method
is a three-component based technique. Though the new technique is
applicable to all distance ranges, it is more advantageous for
regional and local distance range studies. Another important
difference between the PAI-S/K and the new technique is the
application of the normalization scheme that helps to reduce the
variation in the computed values of skewness and kurtosis from
window-to-window.

The third difference, which is a striking one, between the PAI-S/K
and the technique in this paper is a correction procedure, procedure
for making correction to the P-wave arrival time estimate. To the
best of our knowledge, this new procedure is introduced in this
study for the first time to solve such problems. The absolute value
of the ratio of right-hand-side (RHS) to left-hand-side (LHS) HOS
values gives a very good estimate of the accurate arrival
correction. Saragiotis et al. \cite{REF3} PAI-S/K method uses a
maximum slope correction procedure. Figures 6 to 9 show a comparison
between these two approaches and our results indicate clear
improvements when using the absolute values of the ratios of the
adjacent values of HOS (abs(RHS/LHS) of HOS values) as compared to
using the slope (derivative) of the HOS values. On Figures 6 and 7,
the new ratio method for skewness implemented here suggests 11
sample correction while the maximum slope in PAI-S/K method suggests
5 sample correction. On Figures 8 and 9, the ratio approach for
kurtosis suggests 11 sample correction while the slope approach of
PAI-S/K technique indicate 1 sample correction. The actual
correction required can be seen by closely looking at the P-arrival
very closely (Figure 10).

Figure 10 indicates that the required correction is 13 samples. Our
approach gives not only a much better correction in both kurtosis
and skewness cases, but also both skewness and kurtosis give
consistent correction values. Like PAI-S/K we suggest to apply both
skewness and kurtosis together to detect a P phase arrival. The use
of both these statistical quantities instead of just one for
detection will constrain and give a better result than using just
one of these quantities. The number of false alarms also decreases
with the use of the two quantities simultaneously than using just
one of them.

\begin{figure}[htb]
\center{\includegraphics[scale=0.5]{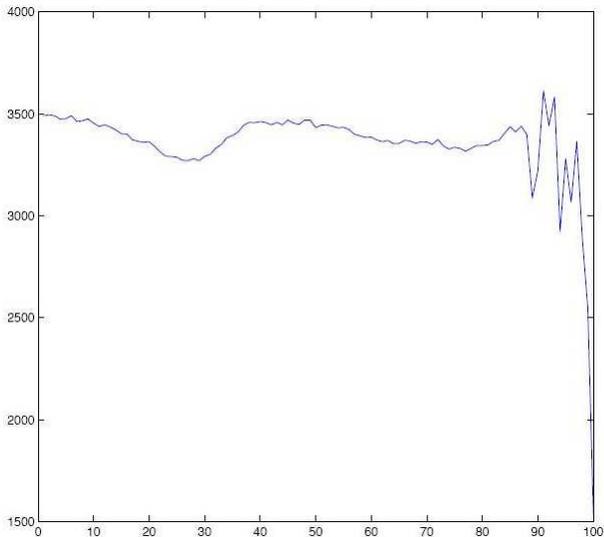}} \caption{Closer
look at the vertical (z) component seismogram including part of the
P-arrival; P-arrived about 12 samples ahead of the end of the
displayed seismogram.} \label{Fig}
\end{figure}

\section{Conclusion}

This article has attempted to make use of vector magnitude of three
component seismograms in order to improve P-wave arrival detection
system. Many single component (usually the vertical component),
single-station based methods have been developed for a P-wave
arrival detection and picking. Since seismic waves from regional
events approach a seismic station (sensor) in a more horizontal
incidence than seismic waves from teleseismic events, the energy is
generally distributed among the horizontal and vertical recording
channels. Thus, it is not only advantageous to develop a method that
makes use of all the three components recorded by the single station
in a combined form but it also gives the technique more generality.
Though the method is applicable to all distance ranges, it is more
advantageous for regional and local distance range studies. We
investigated the application of the normalization to vector
magnitude. In contrast, the HOS values for a P-phase arrival do not
improve as compared to other possible seismic phase arrivals which
may suggest that normalization may play an important role in the
identification of the-more-difficult-to-identify smaller seismic
phases.

We have proposed to apply kurtosis and skewness on the combination
of three component seismograms for picking P wave arrivals
automatically and determined that the method introduced by
Saragiotis et al. \cite{REF3} can be improved if we use the vector
magnitude of seismograms for regional events. This technique makes
use of the information from all the three component data as compared
to the application of the technique on the single component,
vertical, seismogram. The three component seismograms help us
determine the magnitude of the total ground motion through their
vector magnitude. A new approach for making correction of P-wave
arrival time is also proposed and implemented. The absolute value of
the ratio of right-hand-side (RHS) to left-hand-side (LHS) HOS
values gives a very good estimate of the accurate arrival
correction. This has been compared to the derivative (slope) values
correction used in the Saragiotis et al. \cite{REF3} PAI-S/K method
and our results indicate improvements when using abs(RHS/LHS) of HOS
values as compared to using the slope of the HOS values.

We also propose to use both skewness and kurtosis together to detect
a seismic phase arrival. The use of both these statistical
quantities instead of just one for detection will constrain and give
a better result than using just one of these quantities. The number
of false alarms also decreases with the use of thresholds on both
quantities simultaneously than using just one of these two
quantities.

This study has indicated that the RHOS technique can be applied on
vector magnitude of three component seismograms for improved
detection of P-wave arrivals. It is also shown that the new ratio of
HOS values technique can be applied just on the vertical component
seismograms in lieu of the vector magnitude of three component
seismograms and it still gives an improved detection of P-wave
arrivals as compared to that of PAI-S/K. This new procedure also
provides better detection and picking of P-wave arrivals which is
essential for locating earthquake sources more accurately. The
source of the earthquake is the point where slippage between the
fault surfaces or faulting starts inside the earth. Thus, RHOS would
enable us to make more accurate earthquake location using seismic
signals. We strongly believe that this new technique has the
potential to be applied in a similar fashion for accurate detection
and location of fractures in machines or mechanical systems using
acoustic signals.

\end{document}